\documentstyle[prd,aps]{revtex}
\begin{document}
\draft
\twocolumn[\hsize\textwidth\columnwidth\hsize\csname@twocolumnfalse\endcsname
\title{A guided Monte Carlo method for optimization problems}
\author{S. P. Li}
\address{Institute of Physics, Academia Sinica, Taipei, Taiwan 115, R.O.C.}
\date{June, 2001}
\maketitle

\begin{abstract}
We introduce a new Monte Carlo method by incorporating a guided
distribution function to the conventional Monte Carlo method.  In 
this way, the efficiency of Monte Carlo methods is drastically 
improved.  To further speed up the algorithm, we include two more
ingredients into the algorithm.  First, we freeze the sub-patterns  
that have high probability of appearance during the search for optimal 
solution, resulting in a reduction of the
phase space of the problem, a concept inspired by the 
renormalization group equation method in statistical physics.  Second, 
we perform the simulation at a temperature which is within the optimal  
temperature range of the optimization search in our algorithm.  We use this 
algorithm to search for the optimal path of the traveling salesman problem  
and the ground state energy of the Anderson-Edwards spin glass model  
and demonstrate that its performance is comparable
with more elaborate and heuristic methods. 
\end{abstract}

\vspace{2pc}]

Optimization problems arise in areas of science, engineering 
and other fields.  Many of them are NP hard problems, in the sense
that the number of computing steps required to solve the problem 
increases faster than any power of the size of the system.  The
traveling salesman problem (TSP)[1], the spin glass problem[2] 
and the Lennard-Jones
microcluster problem[3] are some of the examples belonging to this class. 
Over the years, people have developed heuristic methods that can 
allow one to possibly obtain optimal solutions in some optimization
problems, e.g. the TSP.  However, these methods are usually specially 
designed  for the 
problem that one is seeking to solve.  In many cases, these heuristic
methods will either not be applicable or will take a much longer 
time to locate the optimal solution of the problem.   
In many practical situations, it is more desirable to obtain
a near-optimal solution quickly than to find the true optimal 
solution of the problem.  Furthermore, it is also preferrable to
have an algorithm that is simple to use and general enough to 
treat various optimization problems.

Since its introduction about half a century ago, the Monte Carlo (MC) method [4]
has been widely used to treat optimization problems.
Despite its simplicity and versatility, it is known that Monte
Carlo method cannot usually give one an acceptable solution since
it can easliy get trapped in local minima of the problem.  In the past 
decades, people have developed new algorithms based on the 
original Monte Carlo method.  Simulated Annealing(SA) [5] is one of the 
algorithms that originates from the Monte Carlo method.  Because
of its simplicity and versatility, SA has been employed to study 
varoius statistical systems with success.  Despite its improvement 
over the original Monte Carlo method, SA also suffers from a similar
problem, that the algorithm can easliy get trapped in local minima.
Progress has been made to improve SA and Monte Carlo methods further
by developling more efficient MC sampling algorithms.  More recent 
developments include algorithms such as the multicanonical method [6] 
and simuulated tempering [7].  
All the above mentioned algorithms are meant to give a more 
efficient MC sampling over the original method.  

In this paper, we introduce a new algorithm by incorporating the Monte
Carlo method with a guided distribution function.  The advantage
of this new algorithm is that it is very simple to use and is 
generally suitable for most optimization problems, whether they
involve continuous or discrete parameter spaces.  The essentials
of this algorithm is from both biology and theoretical physics.  In
the following, we will give the main ingredients of this 
algorithm and apply it to two problems, namely, the traveling
salesman problem and the spin glass problem (SGP).  The task of the TSP
is to find a route through a given set of $N$ cities.  The spin glass
model we are testing here is the three-dimensional Edwards-Anderson 
Ising spin glass model [8] in which Ising spins ($S_i = \pm 1$) are
placed on a three dimensional lattice and there are nearest neighbor
interaction only.  The corresponding Hamiltonian for this model is
$$
H = - \sum_{<i,j>} J_{ij} \sigma_i \sigma_j \, ,
$$
where in our simulation, the $J_{ij}$ takes on values $\pm J$ randomly  
and $\sigma_i , \sigma_j$ can take values +1 or -1.  The task 
here is to find the spin values which minimizes $H$.  As it turns out, 
its performance in these two tests is comparable to the more
elaborate methods and it can obtain near-optimal solutions rather
quickly.

A new substance in this algorithm is the addition of a guided 
distribution function to the original Monte Carlo method.  This
introduction of the guided distribution function is motivated 
from biology.  Similar to the subject of evolutionary 
programming [9], this guided distribution function is based on the 
cooperative effort of individuals.  To begin with, we have
a population of $M$ randomly generated solutions.  We let each of them evolve
independently and according to the conventional Monte Carlo rules, 
i.e. we accept a new solution over its old solution when the new
solution is better fit than the old solution.  If a new solution
is less fit than the old solution, we pick the new solution over
the old solution with respect to a Boltzmann factor.  At this
point, it is equivalent to $M$ individual runs of the Monte Carlo
method.  In each of the $M$ individual runs, we keep record of 
its best fit solution while the individual solution is evolving.  

After a certain number of Monte Carlo steps, a distribution
function of the $M$ best fit Monte Carlo solutions will be recorded.  
Let us use the TSP as an example.  After we
perform a preset number of Monte Carlo steps, each of the 
$M$ independent runs has its own best fit solution, or path.   
In each of these paths, we record the links
between any two cities.  There are a total of $N$ links in each
path and a total of $M N$ links of the $M$ best fit solutions.
One has now a distribution function of the probability of appearance 
among all possible links that connects any two cities.  This
completes our first layer of Monte Carlo simulation.  In our  
discussion, a layer of simulation always means a set of $M$ individual
Monte Carlo runs for a preset Monte Carlo steps plus the evaluation
of the distribution function that we just mentioned.  The distribution  
function that we obtain here will be used as a guided distribution function
in the next layer of Monte Carlo simulation of $M$ individual runs. 

In the next layer of Monte Carlo simulation, a set of $M$ individual
Monte Carlo runs is again performed.  An important point here is that
we also start with $M$ randomly generated solutions
at the beginning of the simulation in this layer.  There is however one more
criterion here---that we pick the links which appear less 
frequent in the guided distribution function and try to change them 
into links with  a higher probability of appearance.  In practice, we
pick one of the two cities that are connected with a less probable link
and connect it to a city where this new link appears more often 
in the guided distribution function.  The new solution will be compared with
the old one to decide whether we keep or disgard it according 
to the conventional Monte Carlo method.  The idea behind this is simple.
The more (less) often that a certain link appears among the $M$ best
fit solutions, the more (less) likely that it would (not) appear in
the optimal solution.  One would then let a given solution 
evolve into solutions with links of high probability of appearance
while avoiding links of low probability of appearance.  The
Monte Carlo method still allows enough fluctuation to search
for optimal, or near-optimal solutions.  After a preset number of
Monte Carlo steps is performed, a new guided distribution function
will again be obtained for the next layer simulation.   
In principle, the more Monte Carlo steps
one performs in each layer, and the more layers of simulations
one has, the better will be the result.

Two more ingredients can be added into this new
algorithm.  The first is the elimination of irrelevant degrees
of freedom.  Again taking the TSP as an example.  There are a total
of $\frac{N(N-1)}{2}$ possible links among the $N$ cities.  Some of
them have a higher probability of appearing in the best fit solutions
while others have very low probability.  We can elminate some
of the links by freezing a link between two cities that exceeds a 
certain probability of appearance in the guided distribution function. 
In practice, after the guided distribution function 
in a layer is obtained, one freezes links
which have a high probability say, 0.8 of appearance before one
performs the next layer simulation.  In this way, one reduces
the available degrees of freedom in the system in the next layer,
thus speeds up the search process.   This idea is inspired by the 
renormalization group equation method in statistical physics 
and is similar in spirit to a recent work [10] which has 
proposed to treat optimization problems using the idea of
renormalization in statistical physics.   
By freezing high probability links, one can obtain
near-optimal solutions within a much faster CPU time.  
One can unfreeze the frozen links during the simulation to avoid 
missing the optimal solution. 

The last ingredient that we include in this algorithm is  
from an observation in a recent work [11].  It was [11] 
observed that the average first passage time to reach the global optimum
of an optimization problem in Monte Carlo simulation was a U-shape
temperature dependent curve.  At the optimal temperature,
the avereage first passage time to find the global optimal will be
the shortest.  This means that for the optimization
problem one has, it is best to perform simulation at the optimal temperature  
so that one can locate the optimal, or near-optimal solution in the
fastest CPU time possible.  The reason for the appearance of such
an optimal temperature can be understood as follows.  
At a given temperature, the system tends to stay 
in states near the equilibrium value determined by that particular
temperature rather than going to the global optimum.  As the system
relaxes toward this equilibrium value, it would fluctuate around this
equilibrium value.  The Gaussian like fluctuation would then carry 
the system from its equilibrium value to the global optimum as long
as there are no insurmountable barriers between them.  When the  
simulation is performed at high temperatures, there are too many
paths available and it would take the system too long to locate the 
path that would reach the global optimum.  As the temperature lowers, 
the number of available paths are reduced.  When the temperature
further decreases, the potential barriers and traps make the system
more difficult to move out of the local optima.  There is 
therefore an optimal temperature at which the two factors
balance each other and the system approaches the global optimum
with the shortest time.   

\vskip 0.2 cm
\noindent
(1) The Traveling Salesman Problem (TSP)

Since we have already used the TSP as an example to explain the
implementation of our algorithm above, we do not need to repeat 
the essential steps here but instead only state the
parameters we used and give some comments at the end.  In the 
simulation, we used $M = 50$ in all the TSP cases that we tested.
We performed 150 Monte Carlo steps in each layer at a 
temperature that is near the optimal temperature in each of the
cases which we give in Table I.  We have tested five cases from
the TSP databank [12]: d198, lin318, pcb442, rat783 and fl1577.   
These cases are selected from the set of sample problem 
instances as listed in [1].  We have performed 100 trial runs in
each case and computed the average value of the best solutions
obtained in each of the 100 trial runs.  We also kept record of
the best solution 
among the 100 trial runs and list them in Table I.  We used
6 layers for the first three cases and 8 layers for the last  
two cases respectively.  To make our discussion simple, we only consider  
2-opt moves in our algorithm.  Other moves can be considered but
is not our main emphasis here.   
\vskip 0.5cm
{\footnotesize{\noindent Table I. Tests on 5 cases from TSPLIB.  
$T$ is the temperature
used for the Monte Carlo simulation, $RQ$([1]=\% above the optimal solution)
 is the relative quality
of the average length of the best paths obtained for the 100
trial runs with respect to the optimal solutions given in TSPLIB, 
$Best$ is the shortest path length obtained in the 100 trial 
runs using the present algorithm and $Op$ is the optimal solution 
given in TSPLIB.}} 
\begin{center}
\begin{tabular}{c  c  c  c  c} \hline \hline  
Case \,\,\,\,\,  & \,\,\,\,\,$T$\,\,\,\,\,  & \,\,\,\,\,$RQ$\,\,\,\,\,  & \,\,\,\,\, $Best$ \,\,\,\,\, & \,\,\,\,\,$Op$ \\ \hline 
d198\,\,\,\,\, & \,\,\,\,\, 10.0\,\,\,\,\, & \,\,\,\,\, 0.0155\%\,\,\,\,\, & \,\,\,\,\, 15780\,\,\,\,\, & \,\,\,\,\, 15780 \\ 
lin318\,\,\,\,\, & \,\,\,\,\, 37.5\,\,\,\,\, & \,\,\,\,\, 0.30\%\,\,\,\,\, & \,\,\,\,\, 42029\,\,\,\,\, & \,\,\,\,\, 42029 \\ 
pcb442\,\,\,\,\, & \,\,\,\,\, 27.5\,\,\,\,\, & \,\,\,\,\, 0.67\%\,\,\,\,\, & \,\,\,\,\, 50798\,\,\,\,\, & \,\,\,\,\, 50778 \\ 
rat783\,\,\,\,\, & \,\,\,\,\, 4.0\,\,\,\,\, & \,\,\,\,\, 0.68\%\,\,\,\,\, & \,\,\,\,\, 8819\,\,\,\,\, & \,\,\,\,\, 8806 \\ 
fl1577\,\,\,\,\, & \,\,\,\,\, 2.75\,\,\,\,\, & \,\,\,\,\, 0.88\%\,\,\,\,\, & \,\,\,\,\, 22273\,\,\,\,\, & \,\,\,\,\, 22249 \\  \hline \hline  
\end{tabular}
\end{center}
\vskip 0.5cm

As can be seen in Table I, the algorithm can locate the optimal
paths in the two cases with fewest cities, i.e. d198 and lin318
while it can obtain near-optimal solutions for the other 
three cases within the number of Monte Carlo steps we preset. 
In general, the more the individual Monte Carlo runs, 
i.e. the larger the $M$, and the more the Monte Carlo steps and layers one
uses, the better will be the solution obtained.  For example, 
in the case of d198, 23 out of 100 trial runs found the optimal 
solution when we perform 150 Monte Carlo steps in each layer.  When 
we increased the Monte Carlo steps in each layer to 200, the average
value of the best runs would now be only 0.00906\% above the optimal 
solution with 41 out of 100 trial runs 
found the optimal solution.  As a comparison, we have 
performed a simulation on the d198 case using the conventional 
Monte Carlo method, for 100 trial runs with 45,000 Monte Carlo 
steps per run.  The number of Monte Carlo steps here is 
equivalent to the number of $M$ (50 here) times the number of 
layers (6 in the case of d198) times the 150 Monte Carlo steps in each layer.
The average value for the best solution of the 100 trial runs
is 1.07\% above the optimal value and the best solution is 15843. 
A further increase of Monte Carlo steps to $10^5$ improves
the average value slightly to 0.90\% above the optimal value.   
This shows the superiority of the present algorithm 
over the conventional Monte Carlo method.  

It is perhaps worth mentioning that locating the optimal or 
near-optimal temperature should help one perform 
simulations in the case of random TSP
in a two-dimensional unit square or something similar.  As an 
example, we have obtained an average optimal temperature around 0.02 
over 10 configurations of 
the random TSP with 100 cities, when 100 Monte Carlo steps are performed
in each layer of a 6 layer simulation.  The optimal temperature range 
is however affected by the number of Monte Carlo steps and the 
moves(we use 2opt moves here).  A rescaling
of the optimal temperature range with respect to the system size, 
i.e. the number of cities, is also needed.   

\vskip 0.2cm
\noindent
(2) The Spin Glass Problem (SGP) 

Spin glasses have been a subject in statistical physics that is
under intensive study.  The simplest spin glass model is the 
Edwards-Anderson model and its Hamiltonian has been given in the
above.  There are exact methods available [13] for such spin glass
models in two-dimensions but are not effective in three-dimensions. 
Estimates of the ground state energy are
available for the three dimensional case [14,15] in the literature for
lattice sizes up to $L = 12$ and could be used for comparison.   
Our task here is to use the present algorithm to estimate the
ground state energy of the three-dimensional Anderson-Edwards 
model within a reasonable amount of CPU time.  
Employing our algorithm to this model is straightforward.  
To begin with, we have a set of $M$ randomly generated spin
configuration for an $L \times L \times L$ lattice with randomly  
assigned $\pm J$ couplings.  We again let each of the configurations evolve
for a preset Monte Carlo steps and record the $M$ best
solutions from the $M$ individual runs.  These $M$ best solutions
are then used to construct the guided distribution function.  
Similar to the TSP, we freeze a sub-spin configuration pattern on this
three-dimensional lattice if it exceeds a certain preset
probability.  In this way, we reduce the spin degrees of freedom
after each layer of simulation.  We again randomly generate $M$
spin configurations at the beginning of each new layer.
There is one thing one needs to keep in mind in constructing the guided 
distribution function of the spin glass problem.  
When all the spins on the three-dimensional lattice changes
sign, the energy of the system does not change.  Therefore, one needs
to fix one of the spins to take on the same value for all the $M$
best solutions when one constructs the guided distribution function.  
For example, we can set the spin value of the
(0,0,0) site of the $L \times L \times L$ lattice to be 1.  If a
particular best solution has a value -1 for this site, then all
the spins have to flip sign in this best solution before one 
records its spin configuration.  

In our simulation, we have used $M = 15$.  We have carried out
simulations for $L$ ranging from 4 to 12 and the result is given in Table II.
Since our purpose here is to obtain the ground state energy of 
the model, we will instead perform several runs for each spin 
configurations and pick the lowest energy as its ground state.
The number of layers and the number of Monte Carlo
steps in each layer that we used will also depend on $L$.  
For example, for $L = 6$, 
we used 2 layers with 150 Monte Carlo steps in each layer and 5
trial runs for each randomly assigned coupling configuration we 
generated.  In the case of 
$L = 8$ sites, we used 6 layers with 200 Monte Carlo steps in each 
layer and 10 trial runs for each randomly assigned coupling 
configuration before we 
take the best solution as the ground state energy for that particular
spin configuration.  The number of Monte Carlo steps and layers 
for different $L$ are given in Table II.  

In the simulation of the spin glass model, we have observed that
unlike the case of the TSP, the temperature range in the Boltzmann 
factor that is most efficient
to locate the ground state or near ground state energy is between  
$T = J$ and $T = 1.1J$ for all the $L$ that we tested and we have
used $T = 1.05J$ for all our simulation of the spin glass model.  
That the optimal temperature range is not sensitive to the
lattice size here is probably related to the fact that one only 
considers the nearest neighbor interaction in the model.    
\vskip 0.5cm
{\footnotesize{\noindent Table II. Tests on $L \times L \times L$ SG instances. 
$m$ is the number of randomly assigned $\pm J$ ($J = 1$ here) cases, 
$MC$ is the number of Monte Carlo steps in each layer and 
$E$ is the average ground state energy of the $M$ spin configurations. 
We have set $J = 1$ in our simulation.  We also include here the
results from Ref[14,15] for comparison.}} 
\begin{center}
\begin{tabular}{c c c c c c} \hline \hline  
$L$ & $m$ & $MC$  & $E$ & Ref[14] & Ref[15] \\ \hline
4 & 200000 & 50  & -1.73750(13) & -1.73749(8) & -1.7370(9) \\  
6 & 50000 & 150  & -1.77128(12) & -1.77130(12) & -1.7723(7) \\ 
8 & 4000 & 200  & -1.77910(28) & -1.77991(22) & -1.7802(5) \\ 
10 & 400 & 400  & -1.7831(8) & -1.78339(27) & -1.7840(4) \\ 
12 & 40 & 600  & -1.7847(14) & -1.78407(121) & -1.7851(4) \\  \hline \hline  
\end{tabular}
\end{center}
\vskip 0.5cm
As shown in Table II, the ground state energy that we obtained
by using our algorithm is comparable with the present estimate of
the ground state energy of the model by using other methods.  If more
layers, Monte Carlo steps and trial runs are used,  
the result will be improved.  

In this paper, we have introduced a new algorithm by
incorporating a guided distribution function into the conventional
Monte Carlo method.  This new algorithm is both biologically and
physically motivated.  By evolving a population of $M$ solutions  
and record their best individual solutions in each layer, we are
relying on the cooperative effort of the population, similar to
many evolutionary algorithms such as genetic algorithms[16]. 
Reducing the computational phase space by eliminating irrelevant
degrees of freedom of the problem as we continue our layer by
layer simulation is similar to the renormalization group equation
method in statistical physics.  Together with the use of an optimal
temperature in Monte Carlo simulation for optimization problems, 
which results in a speed up of optimum search processes, it proves to
be a powerful new algorithm.  It takes about 24 seconds and 9 seconds of  
a Pentium II 450 MHz CPU time to perform 
a trial run of the d198 of the TSP and $L = 8$ of the SGP.  Tests of  
our algorithm in the TSP and SGP are promising.  
Incorporating heuristic methods into this 
algorithm will certainly improve the efficiency and power of its
optimum search.  This algorithm is very general and can be applied
to many optimization problems including discrete optimization 
problems we tested here and optimization problems with continuous 
parameters.  The present algorithm has been applied to the 
Lennard-Jones microcluster problem and can reproduce all the
known ground state energy for the system sizes that the algorithm has
so far tested [17].   We believe that this algorithm should be a simple
and general enough algorithm that can obtain optimal or near-optimal
solutions within a reasonable CPU time for optimization problems.


\begin{references}
\bibitem[\ast]{li}
E-mail address: {\tt spli@phys.sinica.edu.tw}
\bibitem{rei}
G. Reinelt, {\it {The Traveling Salesman: Computational Solutions 
for the TSP Applications}}, (Springer-Verlag, Berlin, 1994). 
\bibitem{you}
See, e.g. A.P. Young (ed.), {\it {Spin Glasses and Random Fields}},
(World Scientific, Singapore, 1998). 
\bibitem{wal}
D.J. Wales, J.P.K. Doye, A. Dullweber and F.Y. Naumkin, The Cambridge
Cluster Database, URL http://brian.ch.cam.ac.uk/CCD.html. 
\bibitem{met}
N. Metropolis, A. Rosenbluth, M. Rosenbluth, A. Teller and E. Teller, 
Journal Chem. PHysics {\bf 21}, 1087 (1953). 
\bibitem{kir}
S. Kirkpatrick, C.D. Jr. Gelatt and M.P. Vecchi, Science {\bf 220}, 
671 (1983). 
\bibitem{ber}
B.A. Berg and T. Neuhaus, Phys. Lett. B {\bf 267}, 249 (1991).  
\bibitem{fri}
E. Marinari and G. Parisi, Europhys. Lett. {\bf 19}, 451 (1992). 
\bibitem{edw}
S.F. Edwards and P.W. Anderson, J. Phys. F {\bf 5}, 965 (1975). 
\bibitem{mic}
See, e.g., Z. Michalewicz, {\it Genetic Algorithms + 
Data Structures = Evolution 
Programs} (Springer-Verlag, New York, New York, 1996). 
\bibitem{hou}
J. Houdayer and O.C. Martin, Phys. Rev. Lett. {\bf 83}, 1030 (1999). 
\bibitem{che}
C.N. Chen, C.I. Chou, C.R. Hwang, J. Kang, T.K. Lee and S.P. Li,
Phys. Rev. E {\bf 60}, 2388 (1999). 
\bibitem{tsp}
http://www.informatik.uni-heidelberg.de/groups
/comopt/software/TSPLIB95/index.html. 
\bibitem{sim}
C. de Simone {\it{et al.,}} J. Stat. Phys. {\bf 84}, 1363 (1996).
\bibitem{pal}
K.F. Pal, Physica A {\bf 223}, 283 (1996). 
\bibitem{har}
A.K. Hartmann, Europhys. Lett. {\bf 40}, 429 (1997). 
\bibitem{gol}
D.E. Goldberg, {\it Genetic Algorithms in Search, Optimization, and Machine
Learning} (Addison-Wesley, Reading, Massachusetts, 1989). 
\bibitem{cho}
C.I. Chou and T.K. Lee (private communication). 
\end{references}
\end{document}